\documentclass[12pt,preprint]{aastex}
\begin{document}
\title{The Accretion Flow - Discrete Ejection Connection in GRS 1915+105}
\author{Brian Punsly\altaffilmark{1}, J\'{e}r$\hat{\mathrm{o}}$me
Rodriguez\altaffilmark{2} and Sergei A. Trushkin \altaffilmark{3}}
\altaffiltext{1}{1415 Granvia Altamira, Palos Verdes Estates CA, USA
90274 and ICRANet, Piazza della Repubblica 10 Pescara 65100, Italy,
brian.punsly1@verizon.net} \altaffiltext{2}{Laboratoire AIM,
CEA/DRF-CNRS-Universit\'{e} Paris Diderot, IRFU SAp, F-91191
Gif-sur-Yvette, France.} \altaffiltext{3}{Special Astrophysical
Observatory RAS, Nizhnij Arkhyz, 369167, Russia and Kazan Federal
University, Kazan, 420008 Russia }
\begin{abstract}
The microquasar GRS~1915+105 is known for its spectacular discrete
ejections. They occur unexpectedly, thus their inception escapes
direct observation. It has been shown that the X-ray flux increases
in the hours leading up to a major ejection. In this article, we
consider the serendipitous interferometric monitoring of a modest
version of a discrete ejection described in Reid et al. (2014) that
would have otherwise escaped detection in daily radio light curves.
The observation begins $\sim 1$ hour after the onset of the
ejection, providing unprecedented accuracy on the estimate of the
ejection time. The astrometric measurements allow us to determine
the time of ejection as $\rm{MJD}\, 56436.274^{+0.016}_{-0.013}$,
i.e., within a precision of 41 minutes (95\% confidence). Just like
larger flares, we find that the X-ray luminosity increases in last 2
- 4 hours preceding ejection. Our finite temporal resolution
indicates that this elevated X-ray flux persists within
$21.8^{+22.6}_{-19.1}$ minutes of the ejection with 95\% confidence,
the highest temporal precision of the X-ray - superluminal ejection
connection to date. This observation provides direct evidence that
the physics that launches major flares occurs on smaller scales as
well (lower radio flux and shorter ejection episodes). The
observation of a X-ray spike prior to a discrete ejection, although
of very modest amplitude suggests that the process linking accretion
behavior to ejection is general from the smallest scales to high
luminosity major superluminal flares.
\end{abstract}


\section{Introduction}
One of the most striking features of Galactic black holes (GBHs) are
the major radio flares. These are large increases in radio flux that
appear to be rapidly moving discrete components in followup radio
interferometric observations \citep{mir94,fen99,dha00,mil05,mil12}.
Major flare ejections (MFEs) of rapidly moving components are rare
states that occur as brief transients in some GBHs. The physics of
the launching mechanism that produces these dramatic events has been
the subject of much speculation \citep{fen04,pun14}. The primary
hurdle to understanding the physics of these powerful events is that
they occur unexpectedly and are quite brief in their inception.
Thus, X-ray telescopes and radio interferometers are never both
pointing at these objects at the precise instance of ejection.
Consequently, the connection between the accretion state (the X-ray
emission) and discrete superluminal ejections has been hampered by
the coarse daily monitoring of these events that evolve on time
scales on the order of hours or minutes. As such, there is not an
agreed upon understanding of the state of the accretion disk at the
time of ejection.
\par In terms of producing
MFEs, GRS~1915+105 is far and away the most prolific. Consequently,
there is a wealth of observational information from monitoring and
pointed observations. Increases in the X-ray light curve have been
associated with MFEs based on daily monitoring \citep{nam06}. Yet,
the coarse time resolution makes it unclear if the X-ray increases
occurred during, after or following the ejection episode. This is a
critical distinction for theorists who are trying to understand the
physical mechanism of the MFEs. The situation is further clouded by
efforts to unify jet phenomena. There is high time resolution
(minutes and seconds) monitoring of oscillatory events that are
discussed in an inclusive context with MFEs, eg. \citep{fen05}. The
notion of X-ray dips preceding superluminal ejections has become a
well accepted notion. Yet, in the study of these oscillatory events,
they are so brief and weak that no apparent motion has ever been
directly detected with a radio interferometer. The notion is so
popular that in the highly publicized work of \citet{mar02} they
claimed that they detected X-ray dips preceding superluminal
ejections in the active galactic nucleus of the Seyfert galaxy 3C120
and this was direct evidence of a commonality with GRS~1915+105.
This is one of the most cited pieces of evidence supporting the
notion of  scale invariance in black hole physics from stellar
massive black holes to supermassive black holes. Our detailed study
of an ejection in GRS~1915+105 aims improve our understanding of the
putative accretion disk - superluminal ejection connection.
\par In a quest to understand the physics of the discrete ejections,
we have been studying the X-ray time evolution immediately preceding
flare production and during the ejections in GRS~1915+105
\citep{pun13,pun15,pun16}. The data being all serendipitous is not
ideal, but we have reached and reaffirmed certain conclusions:
\begin{enumerate}
\item A conspicuous peak in  X-ray flux occurs in the hours preceding the launch of the MFE.
\item During the ejection, the X-ray flux is highly variable.
\item Typically, there is a dip in the X-ray flux
during the ejection, well below the pre-launch flux.
\item The time averaged X-ray flux during the ejection is correlated with the flux preceding
the ejection with a similar, but slightly smaller, magnitude.
\item The X-ray light curve during the MFE often has large local maxima which can exceed the
X-ray flux before the launch. These maxima can occur either during
the ejection or immediately after the ejection episode.
\end{enumerate}
The examination of these findings in the past has been hampered by
crude temporal sampling. MFE ejection times have been estimated from
radio light curves. There is an inherent ambiguity in that optically
thick ejections will not show an increase in the light curve until
they expand to become optically thin at the observing frequency.
Thus, they are not precise. Historically, triggered radio
interferometry initiates $>24$ hrs. after an ejection. Thus,
extrapolating the trajectory of the moving plasmoids back in time
for $>1$ day leads to large uncertainties in the ejection time.
Furthermore, only the strongest MFEs ($> 175$ mJy at 2.3 GHz) have
been studied in the past since one needs a definite strong signal
above the random radio fluctuations of the active source in
GRS~1915+105 in order to determine if one has a certain detection of
a MFE \citep{dha04}.

\begin{figure}
\begin{center}
\includegraphics[width=130 mm, angle= 0]{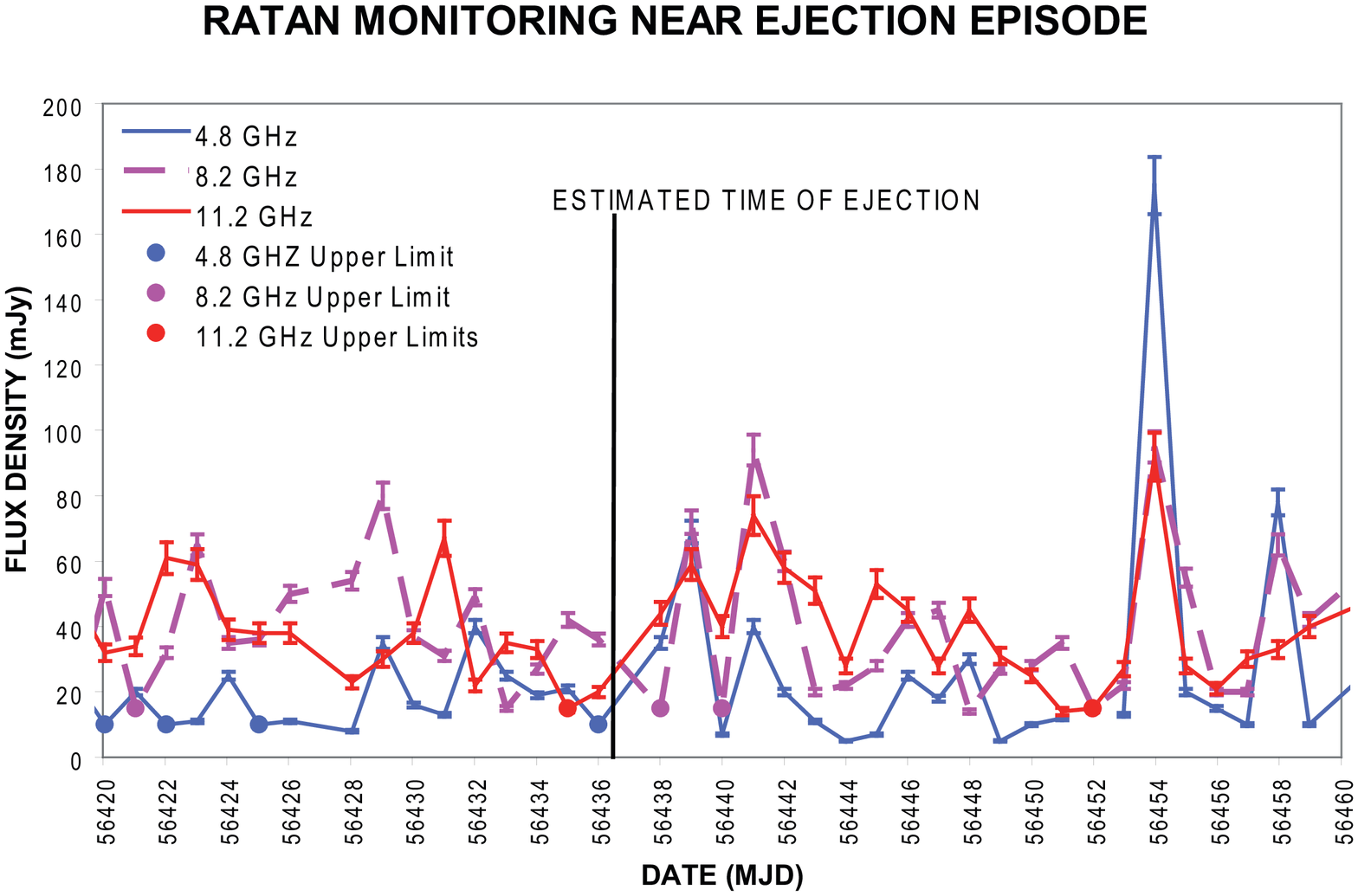}

\end{center}
\caption{The RATAN-600 light curve around the time of the discrete
ejection.}
\end{figure}
\par
We present much more accurate temporal data than has previously been
available of a MFE. We also are able to extend some of the trends
that were noted above to a much weaker ejection. We fortunately have
multi-epoch, snapshot, Very Long Baseline Array (VLBA) observations
of a slightly superluminal ejection with the data sampling beginning
$\sim$1 hour after the onset of the ejection, that were already
published in Reid et al. (2014). We re-analyze here these data
following a slightly different procedure. We adopt the original
strategy to place this VLBA observation in a context of radio
monitoring with the RATAN telescope and confront this radio flare to
contemporaneous X-ray observations  obtained from the MAXI all sky
monitor (http://maxi.riken.jp/mxondem/). The MAXI observatory
typically observes GRS~1915+105 every 1.5 hours and it can
potentially provide useful spectral data in the energy range 1 keV -
20 keV \citep{mat09}. We have downloaded the MAXI data that occurred
within minutes of the ejection as well as two hours before and one
hour after the ejection began. The results of our data analysis are
found in Section 4.

\section{Radio Monitoring} GRS 1915+105 was observed with the RATAN-600
radio telescope at 4.8, 8.2 and 11.2 GHz as part of our 2013
monitoring campaign. The details of these observations have been
described elsewhere \citep{tru08,pun16}. The secondary calibrators
3C286 and PKS1345+12 were used for daily calibration. The temporal
spacing of our monitoring program -- one measurement per day -- is
far too large to estimate the time of the major ejection that is
associated with the radio flare. From \citet{pun13} and Figure 1,
determination of the ejection initiation and end times require
temporal resolution 100 times finer, on the order of an hour or
minutes. However, strong flares are luminous for at least 1 day, so
it is adequate for identifying major ejections \citep{rod99,mil05}.
Figure 1 shows the light curve near the ``suggested" time of the
ejection detected with VLBA on MJD 56436.3 \citep{rei14}. The radio
flux density is low as expected a few hours before the ejection. The
increase in the low frequency flux at 4.8 GHz, 1.7 days after the
ejction, is consistent with the later stages of a discrete ejection.
However, the flux levels are so low for this event and the time
sampling so coarse that the identification cannot be made with
certainty.

\section{The VLBA Observations} GRS~1915+105 was observed with VLBA
at 22.2 GHz on May 24, 2013 (MJD 56436).  The data from the 8
continental VLBA stations were imaged in 7 individual 20 minute
scans. In order to track the motion of the ejection, ``snapshots"
were made from the VLBA observation which have $\sim 20$
interferometer (u,v)-points. With the sparse u-v coverage only peak
fluxes were detectable. One component was detected and its
trajectory is plotted from the data in \citet{rei14} in Figure 2.
The peak flux density at 22.2 GHz was $~6.7 \pm 0.8$ mJy for 6 of
the 7 snapshots. The third (outlier) data point with the larger
error bars was $\lesssim 2$ mJy. The core was not detected and a
conservative upper limit of 3 mJy is placed on its flux density.
\par The trajectory provides information on the the point of
origin for the ejection. There are three things to consider in
establishing the location of the point of origin.
\begin{enumerate}
\item \textbf{Accuracy of Origin Based on Proper Motion}: The expected 22.2 GHz core position based on
the parallax and proper motion fitting results of \citet{rei14}
yield the origin of the plot (which is not necessarily the point of
origin of the ejection). The uncertainty associated with its
location is ($\Delta$RA, $\Delta$Dec) = (0.055 mas, 0.100 mas).
\item \textbf{Optical Depth Effects:} The radio core detected in the parallax measurements
is detected at high frequency (22.2 GHz) and the data has been
checked for a constant flux over the observations in order to
validate the application of a Fourier transformation of the
interferometer data \citep{rei14}. The indication is that the
detections are most likely the optically thick emission commonly
associated with a compact jet \citep{kle02,rus11,gal03}. Thus, one
does not see the point of origin, but the synchrotron self
absorption (SSA) optical depth, $\tau \sim 1$ surface. Modeling the
source of multiple compact jets in GRS~1915+105 in \citet{pun17}
consistently indicated that the peak of the 15 GHz flux density was
$\approx 0.3$ mas from the true position of jet origin. Likewise, we
expect that the peak of the 22.2 GHz emission is offset from the
true point of origin for the putative compact jets detected by
parallax observations. However, being much weaker than the compact
jets considered in the previous models of \citet{pun17}, the
dimensions are likely smaller. Previous jets were $\sim 10$ times
brighter than the putative compact jets monitored for the parallax
measurements \citep{rei14}. The surface area of the $\tau \sim 1$
surface scales with the luminosity (in a highly modeled dependent
manner). Thus, we expect the magnitude of the core shift to be
smaller as well. Furthermore, there is less opacity at higher
frequency and highly simplified models of jets indicate that the
core shifts scales like the observing frequency as $\nu^{-1}$
\citep{bla79}. There are two effects, a smaller dimension in the
weaker compact jets and an upstream shift of the core towards its
true position due to less optical depth at 22.2 GHz. We incorporate
this broad range of possible upstream dislocations by placing the
point of jet origin, corrected for optical depth effects, to be at
$\approx 0.2 \pm 0.1$ mas upstream of the proper motion estimated 22
GHz peak flux density position.
\item \textbf{Statistical Scatter in the Fit to the Trajectory}. The
trajectory is estimated from a least squares fit with uncertainty in
both variables. The data scatter leads to an uncertainty in the best
fit trajectory.
\end{enumerate}
\par Our first step in the determination of the ejection time is to
fit the data by a linear least squares with uncertainty in both
variables \citep{ree89}. Unlike \citet{rei14}, we do not exclude the
outlier point. The best fit to the data scatter in Figure 2 is the
solid line and the dashed lines show the 95\% confidence contours of
the fit. The next step is to find the radio core adjusted for
optical depth effects based on points 1 and 2 above. We displace the
origin 0.2 mas upstream of the point (0, 0) parallel to the
direction given by the slope of the trajectory obtained from our
least squares fit. We find the shifted radio core location based on
astrometry and the shift due to optical depth to be at ($0.15 \pm
0.09 $ mas, $0.13 \pm 0.12$ mas), where the errors are obtained by
adding the errors in points 1 and 2 above in quadrature. This
shifted astrometric core location is noted by the red cross in the
upper right hand corner of Figure 2. The cross represents the 95\%
confidence for the spatial location (i.e., 2$\sigma$) of the shifted
astrometric core. We have two 95\% confidence contours, one for the
astrometric core shifted by $0.2 \pm 0.1$ mas and one for the
trajectory of the ejection. The intersection of these two 95\%
confidence contours provides our 95\% confidence contour for the
point of origin for the ejection shown as the blue triangle. Using
the centroid as the best choice for the point of origin, we find
that it is located at ($0.22^{+0.11}_{-0.22}$ mas,
$0.02^{+0.11}_{-0.13}$ mas) with 95\% confidence.
\par Using this estimate for the point of origin for the jet, we can
compute the time of the ejection by plotting the displacement from
the point of origin versus time and extrapolating backwards in time.
This is done in Figure 3. The least squares fit with uncertainty in
both variables is given by the solid line with the 95\% contours
shown as the dashed lines. The slope of the line is 0.97 mas/hr. The
time of origin from the fit is MJD $56436.274\pm 0.012$. However,
this does not include the uncertainty that is associated with the
point of origin noted at the end of the paragraph above. Using the
speed of the ejection from the slope of the best linear fit to the
observed relationship in Figure 3, we can translate the uncertainty
in location of the point of origin to an uncertainty in time and add
this uncertainty in quadrature with the uncertainty associated with
the fit. We obtain a 95\% confidence contour for the time of
ejection,
\begin{equation}
\rm{EJECTION \; TIME} = \rm{MJD}\, 56436.274^{+0.016}_{-0.013} \;.
\end{equation}

\begin{figure}
\begin{center}
\includegraphics[width=170 mm, angle= 0]{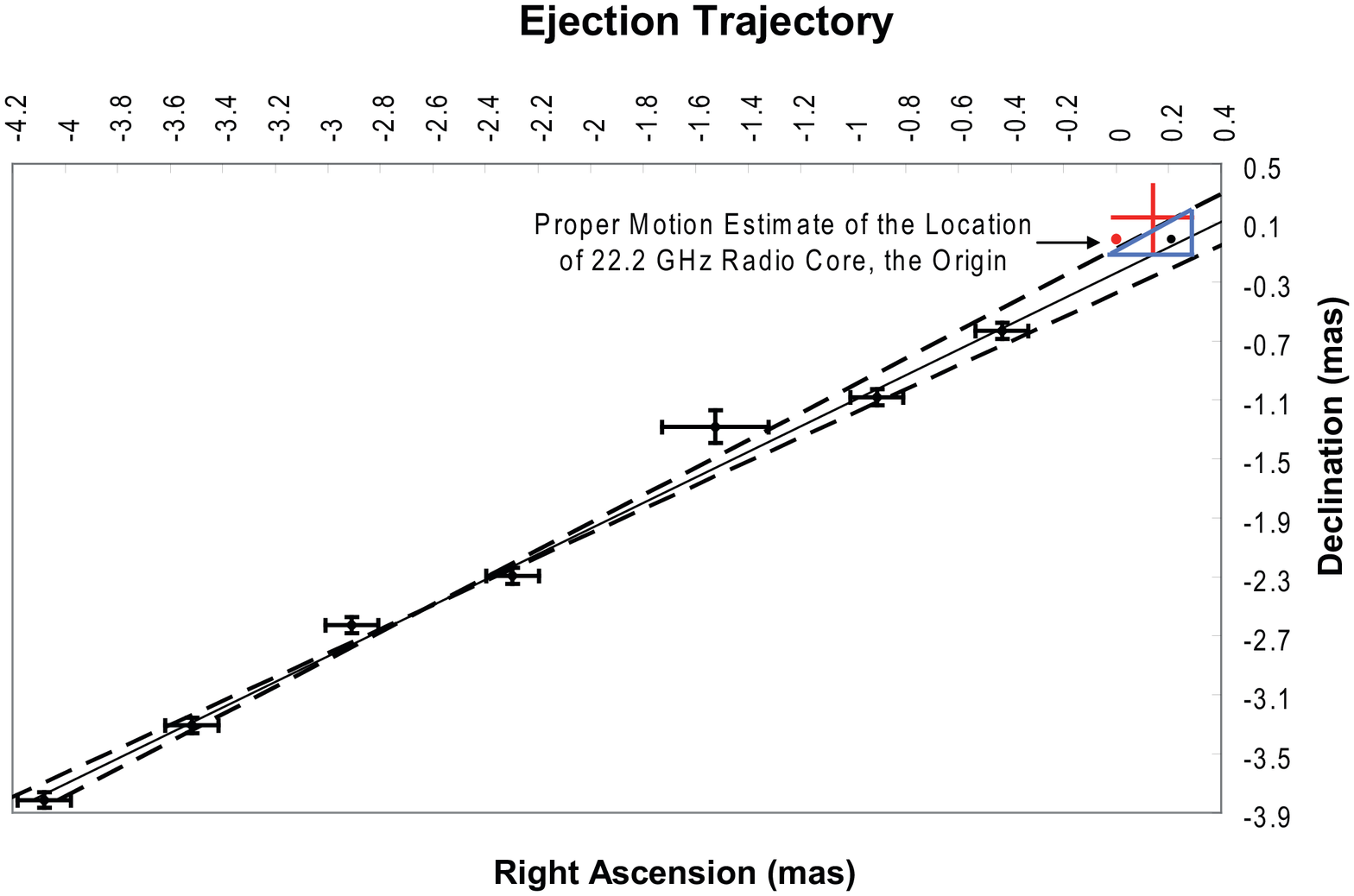}

\end{center}
\caption{The trajectory of the ejection from a series 22.2 GHz VLBA
snapshot images on MJD 56436. The 22.2 GHz radio core (the red dot)
was not detected, but its position is estimated from a series of
proper motion measurements used in the parallax observations of
GRS~1915+105 \citep{rei14}. Optical depth effects shift the true
core position upstream (parallel to the jet trajectory) by an amount
that we estimate from an analysis of 15 GHz VLBA images to be $0.2
\pm 0.1$ mas. The red cross represents the 95\% confidence contour
for this location. The dashed lines are the 95\% confidence contours
for the least squares fit with uncertainty in both variables of the
VLBA determined trajectories. The intersection of the two
independent sets of 95\% confidence contours provides our 95\%
confidence contour for the the point of origin for the ejection that
is bounded by the blue triangle. The black dot, the centroid of the
triangle, is our best estimate of the position of the point of
origin of the discrete ejection.}
\end{figure}
\begin{figure}
\begin{center}

\includegraphics[width=100 mm, angle= 0]{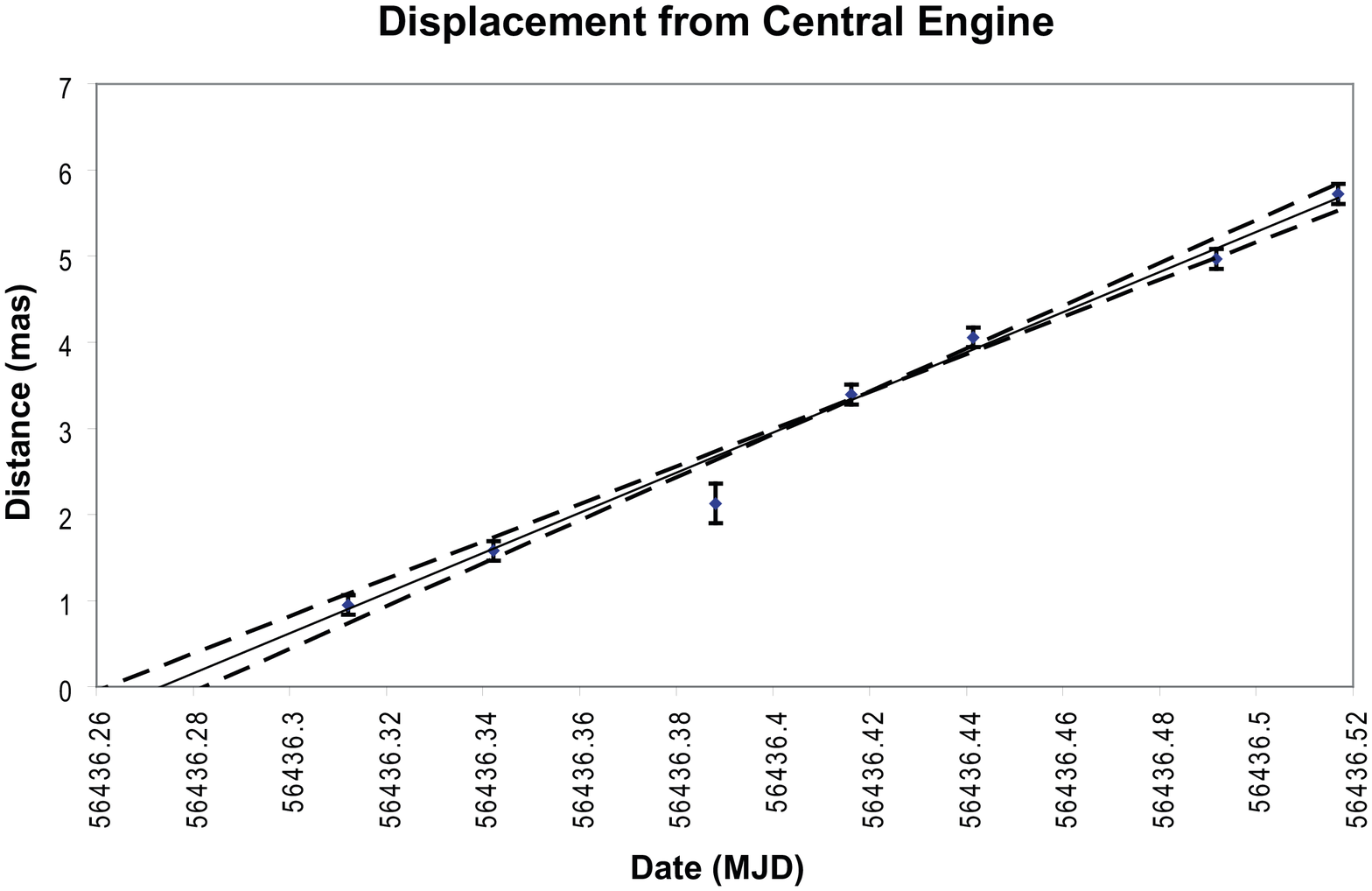}
\end{center}
\caption{The displacement of the discrete ejection from the point of
origin estimated in Figure 2. The data is fit by least squares with
uncertainty in both variables. The uncertainty in the time
measurement (10 minutes) is less than the size of the data point and
uniform.}
\end{figure}
\par Because the precise timing of the X-ray variations and the
ejection time is critical in order to extract the physical mechanism
responsible for discrete ejections, the exercise above is necessary.
The method used by \citet{rei14}, while providing results consistent
with ours, suffers from a high uncertainty on the true core position
leading to slightly different times of ejection when considering the
North-South and East-West ejections. Here by precisely re-setting
the origin of the ejection we strengthen and refine the zero-time of
the ejection. Figure 2 shows that our estimated point of origin for
the jet is $\approx 1 \sigma$ from the estimated radio core
position. This type of accuracy is essential for understanding the
precise timing of the event.

\section{MAXI Observations}
We downloaded the data from the MAXI pipeline around the time of the
ejection.  The light curves of the 2.0 keV - 6.0 keV count rates and
the 2.0 keV -20 keV count rates are plotted in Figure 4. The
ejection time estimated in the last section is indicated by the
dotted black vertical line. The thickness represents the range of
uncertainty. For more than 24 hours GRS~1915+105 is in a low X-ray
state from MJD 56435.047 to MJD 56436.080. Typically, data is taken
every 1.5 hours, but a key data point was missed after this. Thus,
with this limitation, we note that 4 hours later the X-ray flux is
at a maximum. The X-ray luminosity peak occurs on MJD 56436.259.
Comparing this date to Equation (1), this elevated X-ray flux
precedes the onset of the ejection by $21.8^{+22.6}_{-19.1}$ minutes
with 95\% confidence. This is the finest and most accurately
determined juxtaposition in time of an X-ray spike and a MFE. We
have clearly verified point 1 from the Introduction. The flux and
count rate in both the 2.0 keV - 6.0 keV and 2.0 keV - 20.0 keV
windows increases by $\gtrsim 100\%$ in $<4.3$ hrs (Table 1 and
Figure 3). It is uncertain if the ejection is still taking place
during the next MAXI observation on MJD 56436.323.
\par Based on our estimate of the ejection
time above, we use the MAXI spectral data to look for spectral
evolution before and during ejection. The sensitivity of MAXI is low
(low effective area), and we needed to bin the data over the
quiescent period preceding the flare in order to better constrain
the spectral fits. This was acceptable because the source was slowly
varying in time. Our spectral fits are based on an absorbed powerlaw
model ({\tt{tbabs*powerlaw}} in {\tt{XSPEC}} terminology) with
abundances obtained from \citet{wil00}. The absorbed powerlaw model
applied to the MAXI data from 1 keV to 20 keV are described in Table
1. The first column gives the date of the observation. The next
column is the fitted column density of hydrogen, $N_{H}$. The third
column is the photon index of the power law fit to the data,
$\Gamma$. Columns four and five are the absorbed flux and the
unabsorbed (intrinsic) flux that results from removing the absorbing
effects of $N_{H}$, respectively. The last column is the reduced
$\chi^{2}$ of the fit. The errors on the fitted parameters are at
90\% confidence.
\begin{table}
\caption{Parametric Fits to MAXI Data Near the Flare Ejection}
{\footnotesize\begin{tabular}{ccccccc} \tableline\rule{0mm}{3mm}
Date &  $N_{H}$ & $\Gamma$ & Observed Flux  & Intrinsic Flux & Reduced\\
 &   &  & 1 - 10 keV  & 1 - 10 keV & $\chi^{2}$(dof)\\
MJD &  $10^{22} \mathrm{cm}^{-2}$& &  $10^{-8} \mathrm{ergs}/\mathrm{sec-}\mathrm{cm}^{2}$ & $10^{-8} \mathrm{ergs}/\mathrm{sec -} \mathrm{cm}^{2}$ & \\
\tableline \rule{0mm}{3mm}
$56436.05 \pm 0.04$ & $7.2^{+2.3}_{-1.9}$ &$2.4 \pm 0.2$ & $1.0  $ & $2.7 $ & 0.69(26)\\
$56436.18 \pm 0.02$   &  $8.8^{+2.8}_{-2.2}$  & $2.8\pm 0.3 $  & $ 1.5 $ & $5.4 $ & 0.89(20)\\
$56436.26\pm 0.01$   &  $8.1^{+2.3}_{-1.8}$  & $2.7 \pm 0.2$ & $2.0 $ & $6.4 $ & 1.15(25)\\
$56436.65\pm 0.01$   &  $6.2^{+3.5}_{-2.9}$  & $2.4 \pm 0.3 $ & $1.5 $ & $3.7 $& 0.3(20)\\

\end{tabular}}
\end{table}
There are some evident trends in Table 1 for the 56436.27 flare.
Obviously, the luminosity increases before the ejection. But, we
also see $\Gamma$ steepens preceding the flare. This was observed in
MFEs previously observed with MAXI \citep{pun16}. The uncertainty in
the photon index in Table 1 makes this phenomenon less than
statistically significant. We do not see an increase in $N_{H}$
immediately before the flare as previously observed with MAXI
\citep{pun16}. The combination of a weak flare and the low
sensitivity of MAXI makes the spectral analysis difficult and it
might be that for such a weak flare that trends are undetectable
with any significance. We also note a few limitations of the data.
First, the fit to the last data entry in Table 1 is poor. There are
just not enough counts to get good spectral information. Binning or
combining observations is unfortunately not possible here, there is
only the one widely spaced observation (shown in Figure 4) over the
next two days. Furthermore, the changes are so rapid and the fluxes
so low for this flare that it has poor statistics.

\begin{figure}
\begin{center}
\includegraphics[width=130 mm, angle= 0]{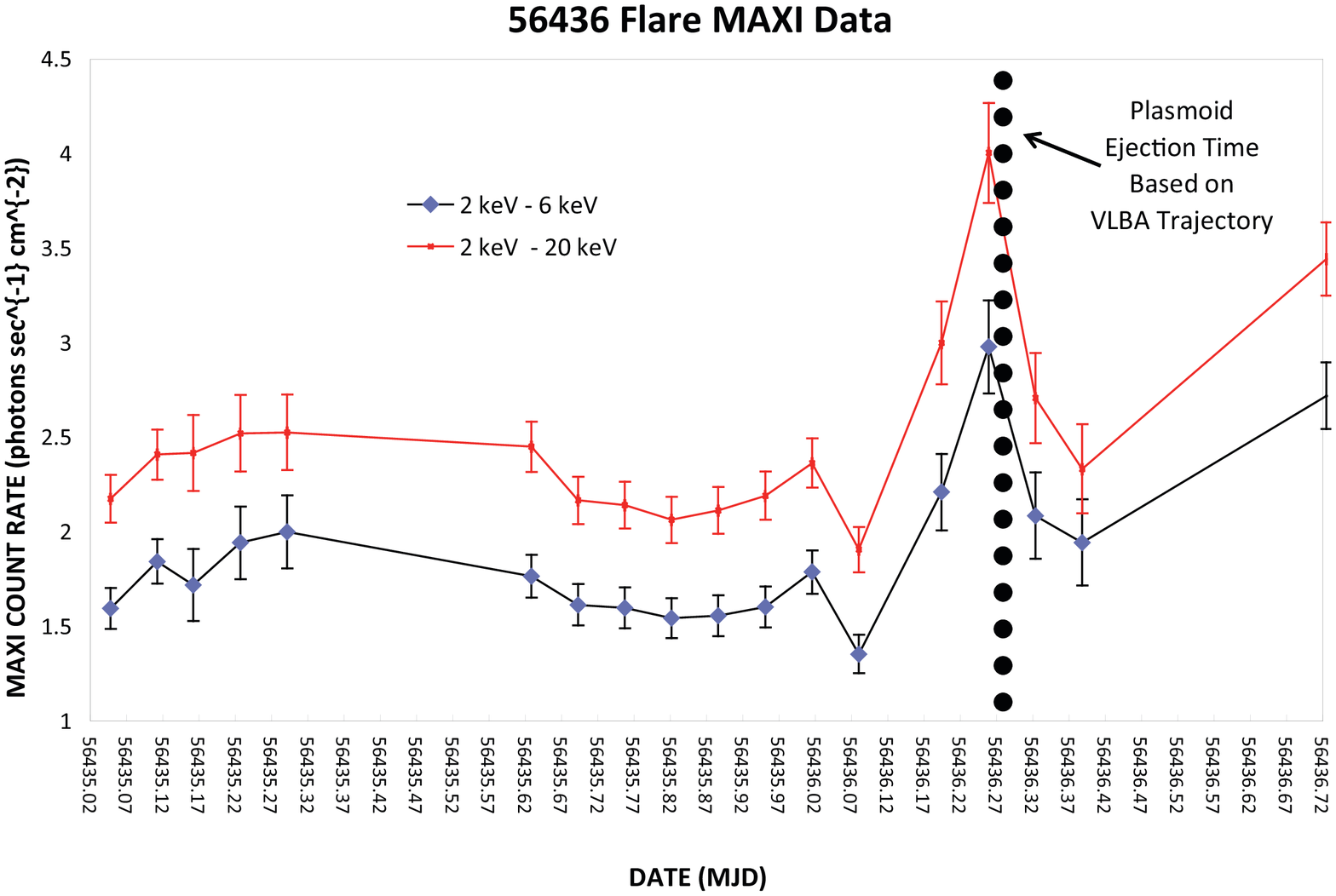}
\end{center}
\caption{The MAXI light curve near the time of the flare. Note the
X-ray flux peak near the time of the ejection.}
\end{figure}
\section{Discussion}
In this article, we re-examined a VLBA observation beginning $\sim
1$ hr after the onset of the ejection of a plasmoid of modest
luminosity in order to explore the disk-jet connection for
superluminal ejections with unprecedented accuracy. We used this
fortunate circumstance in combination with X-ray monitoring with
MAXI to verify that an increase in X-ray luminosity preceded the
ejection. We were able to verify that the X-ray flux begins to
increase 2 - 4 hours before the ejection and the elevated flux
persists within $21.8^{+22.6}_{-19.1}$ minutes of the onset of the
ejection with 95\% confidence. This is one of the five X-ray
properties (listed in the Introduction) detected before in larger
MFEs with coarse time resolution of X-ray coverage and much coarser
estimates of the times of ejection. The other four properties noted
in the Introduction could not be verified, due to the fact that the
ejection was relatively weak and brief and the MAXI time resolution
and sensitivity were not adequate. The most interesting aspect
eluded us due to insufficient time sampling of the X-ray data: when
does the rise in X-ray luminosity end relative to the ejection time?
We performed time resolved spectroscopy with MAXI even though the
count rates were generally small to moderate. We did see a trend
(although not statistically significant with the low number of
counts) previously detected with MAXI that the spectral index of the
X-ray power-law steepens when the X-ray luminosity increases before
ejection \citep{pun16}.
\par Our results suggest a possible link between weak radio ``bubbles" and
MFEs. Firstly, weak radio ``bubbles" with flux densities of 10 mJy -
60 mJy at 15 GHz have been associated with X-ray cycles in X-ray
classes $\nu$, $\lambda$, $\kappa$ and $\beta$ \citep{rod08,rod09}.
X-ray spikes occur on the order of seconds to minutes before an
increase in radio flux occurs. The spikes are accompanied by a
softening of the X-ray spectrum similar to the spectral steepening
seen here and in other MAXI observations of MFEs \citep{pun16}. The
discrete ejection observed here is $\approx 45$ mJy at 8.2 GHz (1.7
days after the ejection according to Figure 1) and forms a possible
example of a common ejection mechanism that straddles the luminosity
gap between the fainter ``bubbles" ejections and the brighter MFEs.
Furthermore, the phenomena of an X-ray spike preceding ejections
seems to extend to stronger radio emission, up to 120 mJy in other
quasi-periodic cycles \citep{pra10}. However, we note that no
discrete ejections were monitored. The unification of these
phenomenon has some significant discrepancies. The single discrete
ejection discussed here and MFEs are temporally different from
quasi-periodic episodes described in
\citep{rod08,rod09,pra10,dha00}. There is no cycle, the event
happened once. Also, the time scale for the X-ray spike to grow is
much larger for the one time discrete ejections than for
quasi-periodic cycles \citep{pra10}. Furthermore, strong
quasi-periodic radio states showed no discrete components in VLBA
monitoring \citep{dha00}. The radio image is morphologically similar
to a blow-torch flame with no concentrations of intensity. This
might be an artifact of blurring from multiple ejections moving a
beam width in $\sim 1$ hour and poor u-v coverage from only 10
antennas.

\begin{acknowledgements}
We would like to thank Mark Reid for generously sharing the details
of his VLBA data reduction. This research has made use of MAXI data
provided by RIKEN, JAXA and the MAXI team. JR acknowledge funding
support from the French Research National Agency: CHAOS project
ANR-12-BS05-0009 (http://www.chaos-project.fr). JR also acknowledges
financial support from the UnivEarthS Labex program of Sorbonne
Paris Cit\'e (ANR-10-LABX-0023 and ANR-11-IDEX-0005-02). SAT is
thankful to Russian Foundation for Base Research for support (Grant
N12-02-00812). We were also very fortunate to have a referee who
offered many useful comments.
\end{acknowledgements}


\begin{thebibliography}{}
\bibitem[Blandford and K{\"o}ingl(1979)]{bla79}Blandford, R. and K{\"o}nigl, A. 1979, ApJ \textbf{232} 34
\bibitem[Dhawan et al.(2000)]{dha00}Dhawan, V., Mirabel, I.F., Rodriguez, L. 2000, ApJ 343 373
\bibitem[Dhawan et al(2004)]{dha04}Dwahan, V., Muno, M., Remillard, R. 2004, in \emph{X-Ray Timining 2003 Rossie
and Beyond AIP Conference Proceedings}, Volume 714, pp.150-153
(2004)
\bibitem[Fender et al.(1999)]{fen99}Fender, R. et al., 1999, MNRAS 304 865
\bibitem[Fender et al.(2004)]{fen04}Fender, R., Belloni, T., Gallo, E. 2004, MNRAS 355 1105
\bibitem[Fender and Belloni(2004)]{fen05}Fender, R., Belloni, T. 2004, ARA\&A, 42, 317
\bibitem[Gallo et al.(2003)]{gal03}Gallo, E., Fender, R., Pooley, G. 2003, MNRAS 344 60\
\bibitem[Klein-Wolt et al(2002)]{kle02}Klein-Wolt et al., 2002, MNRAS 331 745
\bibitem[Marscher et al.(2002)]{mar02} Marscher, A., Jorstad, S., G´omez, J.-L., Aller, M. F.,
Ter¨asranta, H., Lister, M., Stirling, A., 2002, Nature 417, 625
\bibitem[Matsuoka et al.(2009)]{mat09}Matsuoka, M. et al.., 2009, PASJ, 61, 999
\bibitem[Miller-Jones et al.(2005)]{mil05}Miller-Jones, J. et al. 2005, MNRAS 363 867
\bibitem[Miller-Jones et al.(2012)]{mil12}Miller-Jones, J. et al. 2012, MNRAS 421
468
\bibitem[Mirabel and Rodriguez(1994)]{mir94}Mirabel, I.F., Rodriguez, L. 1994, Nature 371 46
\bibitem[Namiki et al (2006)]{nam06} Namiki, M., Trushkin, S., Kotani, T., Kawai, N., Bursov, N., Fabrika, S.  206,
VI Microquasars Workshop: Microquasars and Beyond, September 18-22,
2006, Societa del Casino, Como, Italy PoS(MQW96)083
\bibitem[Prat et al(2010)]{pra10}Prat, L., Rodriguez, J, and Pooley, G. 2010, ApJ 717 1222
\bibitem[Punsly and Rodriguez(2013a)]{pun13} Punsly, B., Rodriguez J. 2013a, ApJ 764 173
\bibitem[Punsly and Rodriguez(2013b)]{pun14} Punsly, B., Rodriguez J. 2013b, ApJ 770 99
\bibitem[Punsly and Rodriguez(2013c)]{pun15} Punsly, B., Rodriguez J. 2013c, MNRAS 435 2322
\bibitem[Punsly and Rodriguez(2016)]{pun17} Punsly, B., Rodriguez J. 2016,
to appear in ApJ http://arxiv.org/abs/1603.07675
\bibitem[Punsly Rodriguez and Trushkin (2014)]{pun16} Punsly, B., Rodriguez J., Trushkin< S. 2014, ApJ 783
133
\bibitem[Reed(1989)]{ree89} Reed, B. 1989, Am. J. Phys. 57 642
\bibitem[Reid et al.(2014)]{rei14} Reid, M. et al. 2014, ApJ 796 2
\bibitem[Rodriguez and Mirabel(1999)]{rod99} Rodriguez, L., Mirabel, I.F. 1999, ApJ 511 398
\bibitem[Rodriguez et al(2008a)]{rod08} Rodriguez, J. et al 2008, ApJ \textbf{675}
1436
\bibitem[Rodriguez et al(2008b)]{rod09} Rodriguez, J. et al 2008, ApJ \textbf{675}
1449
\bibitem[Rushton et al.(2010a)]{rus10}Rushton, A., Spencer, R. E., Pooley, G., and Trushkin, S. 2010, MNRAS 401
2611
\bibitem[Rushton et al.(2010b)]{rus11}Rushton, A., Spencer, E., Fender, R. and Pooley, G. 2010, A \& A 524 29
\bibitem[Trushkin et al.(2008)]{tru08} Trushkin, S. A., Bursov, N.N., Nizhelskij N.A.
2008, AIP Conference Proceedings,  1053, 219.
\bibitem[Wilms et al.(2000)]{wil00}Wilms, J., Allen, A., McCray, R. 2000, ApJ 524 914
\end{thebibliography}
\end{document}